\documentclass[aps,pra,reprint,superscriptaddress,showpacs,twocolumn]{revtex4}

\usepackage{amsfonts}
\usepackage{amsmath}
\usepackage{amssymb}
\usepackage{graphicx}
\usepackage{color}

\begin{document}

\newcommand{\ket} [1] {\vert #1 \rangle}
\newcommand{\bra} [1] {\langle #1 \vert}
\newcommand{\braket}[2]{\langle #1 | #2 \rangle}
\newcommand{\proj}[1]{\ket{#1}\bra{#1}}
\newcommand{\mean}[1]{\langle #1 \rangle}
\newcommand{\opnorm}[1]{|\!|\!|#1|\!|\!|_2}
\title{Enhancing quantum entanglement by photon addition and subtraction}


\author{Carlos Navarrete-Benlloch}\email{carlos.navarrete@mpq.mpg.de}
\affiliation{Departament d'\`Optica, Universitat de Val\`encia, Dr. Moliner 50, 46100 Burjassot, Spain}
\affiliation{Max-Planck-Institut f\"ur Quantenoptik, Hans-Kopfermann-Strasse 1, 85748 Garching, Germany}
\affiliation{Research Laboratory of Electronics, Massachusetts Institute of Technology, Cambridge, Massachusetts 02139, USA}

\author{Ra\'ul Garc\'ia-Patr\'on}
\affiliation{Max-Planck-Institut f\"ur Quantenoptik, Hans-Kopfermann-Strasse 1, 85748 Garching, Germany}
\affiliation{Research Laboratory of Electronics, Massachusetts Institute of Technology, Cambridge, Massachusetts 02139, USA}

\author{Jeffrey H. Shapiro}
\affiliation{Research Laboratory of Electronics, Massachusetts Institute of Technology, Cambridge, Massachusetts 02139, USA}

\author{Nicolas J. Cerf}
\affiliation{QuIC, Ecole Polytechnique de Bruxelles, CP 165, Universit\'e Libre de Bruxelles, 1050 Brussels, Belgium}
\affiliation{Research Laboratory of Electronics, Massachusetts Institute of Technology, Cambridge, Massachusetts 02139, USA}

\date{\today}

\begin{abstract}
The non-Gaussian operations effected by adding or subtracting a photon on the entangled
optical beams emerging from a parametric down-conversion process
have been suggested to enhance entanglement.
Heralded photon addition or subtraction is, as a matter of fact, at the heart
of continuous-variable entanglement distillation.
The use of such processes has recently been experimentally demonstrated
in the context of the generation of optical coherent-state
superpositions or the verification of the canonical commutation relations.
Here, we carry out a systematic study of the effect of local photon additions or subtractions on a two-mode squeezed vacuum state, showing that the entanglement generally increases with the number of such operations. This is analytically proven when additions or subtractions are restricted to one mode only, while we observe that the highest entanglement is achieved when these operations are equally shared between the two modes. We also note that adding photons typically provides a stronger entanglement enhancement than subtracting photons, while photon subtraction performs better in terms of energy efficiency. Furthermore, we analyze the interplay between entanglement and non-Gaussianity, showing that it is more subtle than previously expected.
\end{abstract}



\maketitle

\section{Introduction}

Quantum information processing with Gaussian continuous variables is
a well-established subfield of quantum information sciences today, see e.g. \cite{Cerf-book,Braunstein05,WeedbrookUN}.
Quantum key distribution, for example, can be carried out dealing with Gaussian states and measurements
only \cite{Cerf2001,Grosshans2003,Cerf-Grangier-2007}. Nevertheless, non-Gaussian quantum states and operations
are indispensable to perform certain other continuous-variable quantum information tasks,
such as quantum entanglement distillation \cite{Eisert02,Fiurasek02,Giedke02},
quantum error correction \cite{Niset2009}, or universal quantum computation \cite{Lloyd99}.
In addition, any Bell test of quantum nonlocality that relies on Gaussian measurements only necessarily requires the preparation of a
non-Gaussian entangled state \cite{Wenger2003,Carmichael2004,Garcia2004,Garcia2005,Jeong2008}, while a quantum bit commitment protocol
that is secure against Gaussian cheating must necessarily involve a non-Gaussian resource state \cite{Magnin2010}.

\begin{figure*}[t]
\includegraphics[width=0.9\textwidth]{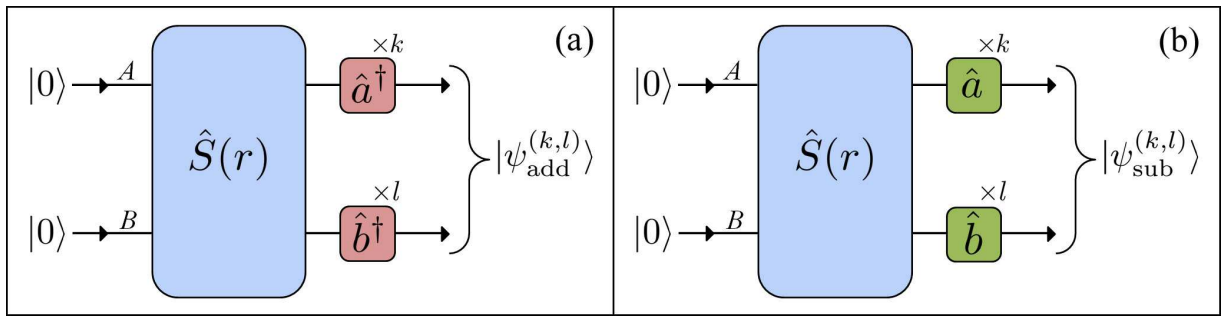}
\caption{\label{Fig1}(Color online) Degaussification schemes analyzed in this work. Alice and Bob share a two-mode squeezed vacuum state, which
results from applying a two-mode squeezing operation $\hat{S}(r)$ on two vacuum modes ($A$ and $B$).
They perform $k$ and $l$ photon additions (a) or subtractions (b) on their corresponding modes,
generating the output states $|\psi_{\mathrm{add}}^{(k,l)}\rangle$
or $|\psi_{\mathrm{sub}}^{(k,l)}\rangle$, respectively.}
\end{figure*}

Deterministically producing a non-Gaussian quantum optical state by using the Kerr effect is unfortunately unfeasible today because 
it requires quite high (called giant) optical nonlinearities, which are not accessible in the laboratory. Probabilistic degaussification schemes, 
however, have been shown feasible based on photon addition and subtraction. Not being unitary, photon addition or subtraction can only be achieved 
probabilistically, that is, conditioned on a particular measurement outcome. One thus refers to it as {\it heralded} photon addition or subtraction. 
The effect of photon subtraction can be obtained by sending a small fraction of the optical beam on an avalanche photodiode and conditioning the use of 
the remaining fraction of beam upon photon-counting events \cite{Opatrny2000,Wenger04}. Photon addition can be achieved as the result of a single-photon excitation 
of the light field produced by parametric down-conversion in a nonlinear medium, conditioning the use of the signal output mode on the detection of a photon in the idler mode \cite{Zavatta04}.

In principle, an arbitrary single-mode state can be prepared by applying a sequence of photon additions  \cite{addition} or subtractions \cite{subtraction} 
that are interleaved with displacement operations; similarly, an arbitrary operation dependending only on the photon number operator can be generated by using appropriate superpositions of addition and subtraction \cite{Fiurasek09}.
On the experimental side, the use of photon subtraction from a squeezed vacuum state has been demonstrated in 
\cite{Polzik,Grangier1,Grangier2,Furusawa1,Furusawa2,Neergaard10} in order to generate low-amplitude 
coherent-state superpositions (Schr\"odinger kitten states) of traveling light (schemes that can be further developed 
to generate larger-size squeezed Schr\"odinger cat states \cite{Marek08}), while the use of photon addition combined with displacements has allowed to 
generate arbitrary superpositions of the first three Fock states \cite{Bimbard10}. Moreover, the ability to superpose different sequences 
of additions and subtractions \cite{Kim08} has enabled checking of the canonical commutation relations \cite{Parigi07,Zavatta09}.

In this paper, we focus on the enhancement of quantum entanglement that results from adding or subtracting an arbitrary number of photons
on the two beams emerging from a nondegenerate parametric down-conversion process, in an attempt to understand the generally 
admitted---but not systematically analyzed---notion that degaussifying the down-converted beams makes them more entangled 
\cite{Opatrny2000,Cochrane02,Olivares03,Kitagawa06,Ourjoumtsev07,Dellano07,Adesso09,Takahashi10,Zhang10,Lee11a}. 
In Section II, we provide the basic equations describing the state obtained by adding $k$ and $l$ photons on 
the first and second beam of the two-mode squeezed vacuum state, respectively, or when similarly subtracting photons. 
Section III is focused on the case $l=0$, where we can analytically prove that quantum entanglement is a monotonically 
increasing function of $k$. Section IV treats the general case ($k,l>0$), and presents an exhaustive analysis of the 
behavior of entanglement enhancement as a function of $k$ and $l$. Note that a subclass of these states ($k=l$) has already been analyzed 
in \cite{Adesso09,Zhang10}. In Sec. V, we investigate a measure of the non-Gaussianity of these photon-added and photon-subtracted states, 
and show that the link between this measure and entanglement is rather subtle. Section VI is devoted to our conclusions.

We would finally like to remark that, even though we are focussing on addition and subtraction of photons in an optical mode, our analysis applies also to addition and subtraction of excitations in other platforms such as phonons in mechanical oscillators \cite{Vanner12}, or polaritons in atomic ensembles \cite{Datta12}. In particular, atomic ensembles might offer several advantages over photonic systems \cite{Hammerer10,Muschik11}: first, they act as a quantum memory, and hence, the state can be re-used if the addition or subtraction protocol does not succeed; secondly, generating the two-mode squeezing interaction between an optical mode and the ensemble is not more challenging from the experimental point of view than generating the beam splitter interaction, which means that addition and subtraction are on an equal footing in the laboratory.





\section{Adding or subtracting photons on a two-mode squeezed vacuum state}

Our starting point is the two-mode squeezed vacuum (TMSV) state. Let us call \textit{A} and
\textit{B} (for Alice and Bob) the involved modes, whose boson operators are respectively
denoted by $\{\hat{a},\hat{a}^\dagger\}$ and $\{\hat{b},\hat{b}^\dagger\}$.
These boson operators satisfy the  usual canonical commutation relations
$[\hat{a},\hat{a}^\dagger]=[\hat{b},\hat{b}^\dagger]=1$.
The TMSV state is obtained by applying the joint operation
$\hat{S}(r)=\exp[r(\hat{a}\hat{b}-\hat{a}^\dagger\hat{b}^\dagger)/2]$ with $r\in[0,+\infty[$,
known as the two-mode squeezer, to the vacuum state of modes \textit{A} and \textit{B}, that is,
\begin{equation}
|\mathrm{TMSV}\rangle = \hat{S}(r)|0,0\rangle = \sqrt{1-\lambda^2}\sum_{n=0}^{\infty}\,
\lambda^{n} |n,n\rangle, \label{TMSV}
\end{equation}
where $\lambda=\tanh r\in[0,1[$, and $|m,n\rangle = |m\rangle_A \otimes |n\rangle_B$.
Here, $\{|n\rangle\}_{n\in\mathbb{N}}$ denotes the number states defined by
$\hat{a}|n\rangle_A = \sqrt{n}|n-1\rangle_A$ and $\hat{a}^\dagger|n\rangle_A = \sqrt{n+1}|n+1\rangle_A$
for mode $A$, and  analogous expressions for mode \textit{B}.

Our aim now is to study how photon addition and subtraction affects the TMSV state when applied locally by Alice and Bob. To this end, we consider the schemes depicted in Fig. \ref{Fig1}. In the first scheme (Fig. \ref{Fig1}a), Alice and Bob add, respectively, $k$ and $l$ photons to their mode.
It is straightforward to show that the final, properly normalized state can be written as
\begin{equation}
|\psi_{\mathrm{add}}^{(k,l)}\rangle = \sum_{n=0}^{\infty} \sqrt{p_n^{(k,l)}} |n+k,n+l\rangle, \label{Add}
\end{equation}
with
\begin{equation}
p_n^{(k,l)} = \frac{\lambda^{2n}}{_2F_1(k+1,l+1;1;\lambda^2)} \binom{n+k}{k} \binom{n+l}{l},
\end{equation}
where $_{2}F_{1}(a,b;c;z)$ is the Gauss hypergeometric function defined as a series expansion
\begin{equation}
_2F_1(a,b;c;z)=1 + \frac{ab}{1!c}z + \frac{a(a+1)b(b+1)}{2!c(c+1)}z^{2} + \cdots
\end{equation}
In the second scheme (Fig. \ref{Fig1}b), photon addition is replaced by photon subtraction, and the output state can be written as
\begin{equation}
|\psi_{\mathrm{sub}}^{(k,l)}\rangle = \sum_{n=k}^{\infty} \sqrt{q_n^{(k,l)}} |n-k,n-l\rangle, \label{Sub}
\end{equation}
with
\begin{equation}
q_n^{(k,l)} = \frac{\lambda^{2(n-k)}}{_2F_1(k+1,k+1;1+k-l;\lambda^2)} \times \frac{ \binom{n}{k} \binom{n}{l}}{\binom{k}{l}},
\end{equation}
where we have assumed that $k\geq l$ (exactly the same expression but interchanging $k$ and $l$ holds for $k<l$).

Note that here we treat photon addition and subtraction as ideal $(\hat{a}^\dagger,\hat{a})$ operations. 
In realistic schemes based on the beam splitter (for subtraction) or the two-mode squeezer (for addition) interaction with an auxiliary vacuum mode, 
this is an approximation which becomes exact only in the unphysical limit of vanishing interaction (e.g, for perfect transmissivity of the beam splitter). 
Nevertheless, as long as the interaction is kept very weak---which then makes successful subtraction or addition events rare, but still frequent enough for applications---the idealized description is a good approximation \cite{Kim08}. In any case, we refer the reader to \cite{subtraction,Marek08,Fiurasek09,Zhang10} 
for a rigorous treatment of photon addition and subtraction in experimentally realistic conditions.

In the following, we analyze the entanglement of these states as a function of the number of photon additions or subtractions. Being pure bipartite states, their entanglement is uniquely measured by the entanglement entropy \cite{Bennett96}, defined for an arbitrary state $|\psi\rangle$ as $E[|\psi\rangle]=\mathcal{S}[\mathrm{tr}_B\{|\psi\rangle\langle\psi|\}]$, where $\mathcal{S}[\hat{\rho}]=-\mathrm{tr}\{\hat{\rho}\log\hat{\rho}\}$ denotes the usual von Neumann entropy of the density operator $\hat{\rho}$. In our case, evaluating this quantity is straightforward because
Eqs.~(\ref{Add}) and (\ref{Sub}) are in Schmidt form, so that  the entanglement entropy of the states
 $|\psi_{\mathrm{add}}^{(k,l)}\rangle$ and $|\psi_{\mathrm{sub}}^{(k,l)}\rangle$ is
\begin{equation}
E_{\mathrm{add}}^{(k,l)} = -\sum_{n=0}^\infty p_n^{(k,l)} \log p_n^{(k,l)}, \label{Eadd}
\end{equation}
and
\begin{equation}
E_{\mathrm{sub}}^{(k,l)} = -\sum_{n=\mathrm{max}\{k,l\}}^\infty q_n^{(k,l)} \log q_n^{(k,l)}, \label{Esub}
\end{equation}
respectively. Unfortunately, we have not been able to carry out these sums analytically except in the trivial case $k=l=0$, where we get
the well-known expression for the entanglement entropy of the TMSV state \cite{WeedbrookUN},
\begin{equation}
E_{\mathrm{TMSV}}(\lambda) = \frac{\lambda^2}{1-\lambda^2} \log \left(\frac{1}{\lambda^2}\right) + \log \left(\frac{1}{1-\lambda^2}\right),
\end{equation}
which is a monotonically increasing function of $\lambda$.
Despite this absence of a closed expression for Eqs. (\ref{Eadd}) and (\ref{Esub}), we are able to analytically derive their dependence
on the parameter $k$ for $l=0$ in the next section.

\section{Entanglement properties for one-mode operations}

\begin{figure}[t]
\includegraphics[width=\columnwidth]{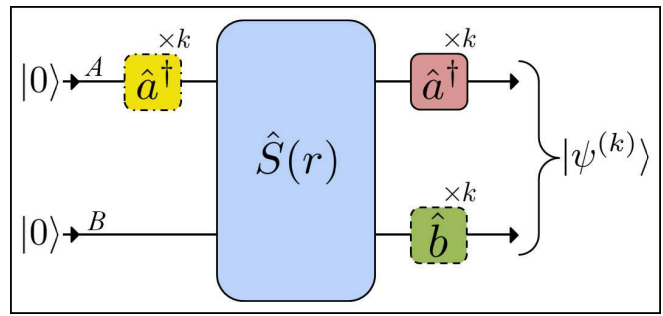}
\caption{\label{Fig2}(Color online) Three equivalent degaussification schemes. Starting with the vacuum state,
each of these three single-mode operations results in the same output state $|\psi^{(k)}\rangle$
up to a normalization factor (i.e., the probabilities of success are different):
Alice adding $k$ photons after the two-mode squeezer $\hat{S}(r)$ (solid-border pink box),
Bob subtracting $k$ photons after $\hat{S}(r)$ (dashed-border green box),
or Alice adding $k$ photons before $\hat{S}(r)$ (yellow box with dashed-dotted border).}
\end{figure}

Let us restrict to the case in which only one of the modes undergoes photon addition or subtraction operations, while the other is unchanged.
In that case, we will be able to prove analytically that the entanglement entropy, either $E_{\mathrm{add}}^{(k,0)}$ or $E_{\mathrm{sub}}^{(k,0)}$,
can only increase with the number of operations $k$. The first thing to note is that the three schemes shown in Fig. \ref{Fig2} lead to
the exact same state
\begin{equation}
|\psi^{(k)}\rangle = \sum_{n=0}^\infty \sqrt{p_n^{(k)}}|n+k,n\rangle,  \label{psi_k}
\end{equation}
with
\begin{equation}
p_n^{(k)} = (1-\lambda^2)^{k+1} \lambda^{2n} \binom{n+k}{n}.
\end{equation}
Note that this state follows from (\ref{Add}) by putting $l=0$ and using
$_2F_1(k+1,1;1;\lambda^2)=(1-\lambda^2)^{-k-1}$.
Using Eq. (\ref{TMSV}), it is easy to prove that
\begin{equation}
\hat{b}|\mathrm{TMSV}\rangle = \lambda \hat{a}^\dagger |\mathrm{TMSV}\rangle, \label{Prop1}
\end{equation}
implying that Alice adding $k$ photons on the first mode (solid-border pink box in Fig. \ref{Fig2})
has the same effect as Bobs subtracting them from the second mode (dashed-border green box in Fig. \ref{Fig2}),
up to a normalization factor related to the success probability of the corresponding operation.
Secondly, it is also easy to check that
\begin{eqnarray}
\hat{S}(r) \left(\hat{a}^\dagger\right)^k |0,0\rangle &=& [S^\dagger(-r) \hat{a}^\dagger \hat{S}(-r)]^k \hat{S}(r) |0,0\rangle \nonumber \\
&=& \frac{1}{\cosh^k r} \left(\hat{a}^\dagger\right)^k \hat{S}(r) |0,0\rangle,
\end{eqnarray}
where we have used Eq.~(\ref{Prop1}) as well as the relation $\hat{S}^\dagger(-r) \hat{a}^\dagger \hat{S}(-r)=\hat{a}^\dagger \cosh r-\hat{b}\sinh r$. Hence, adding $k$ photons before or after the two-mode squeezer is equivalent, except for a normalization factor again related to the probability of success of the non-unitary operation.


\begin{figure*}[t]
\includegraphics[width=0.95\textwidth]{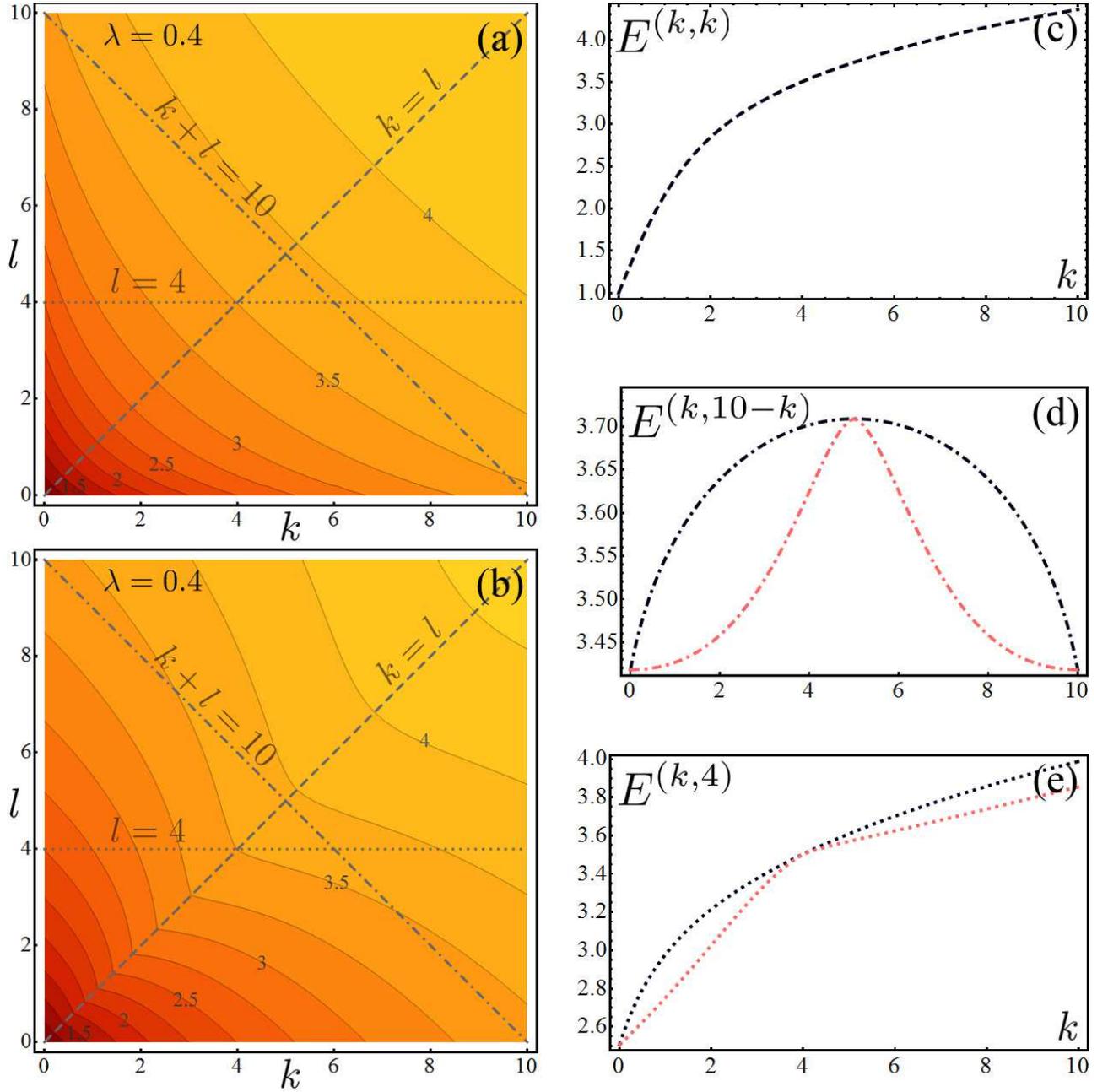}
\caption{\label{Fig3}(Color online) Entanglement entropy of the photon-added and -subtracted states,
$|\psi_\mathrm{add}^{(k,l)}\rangle$ and $|\psi_\mathrm{sub}^{(k,l)}\rangle$,
as a function of the number of operations $(k,l)$ for $\lambda=0.4$. Note that
we have normalized it to the entanglement of the corresponding TMSV state, that is,
 $E_{\mathrm{add}}^{(0,0)}=E_{\mathrm{sub}}^{(0,0)}$ is set to 1. Panels (a) and (b) are
 density plots of $E_{\mathrm{add}}^{(k,l)}$ and $E_{\mathrm{sub}}^{(k,l)}$,
respectively, in the $(k,l)$ space; darker regions correspond to lower entanglement.
The thin lines are the isolines (lines of equal entanglement), while the thick
straight lines correspond to $k=l$ (dashed), $k+l=10$ (dashed-dotted), and $l=4$ (dotted). The
entanglement of the latter three lines is plotted in (c), (d), and (e), respectively, with dark-blue and light-red denoting
the photon-added and photon-subtracted states, respectively. Even though $k$ and $l$ are physically
discrete variables, we have taken them to be continuous variables by using the Gamma function as
an analytic extension of the factorial function (this holds for the rest of the figures in the article).
}
\end{figure*}

Since Eq.~(\ref{psi_k}) is in the Schmidt form, the entanglement entropy of state
$|\psi^{(k)}\rangle$ is easily evaluated as
\begin{equation}
E^{(k)}=-\sum_{n=0}^\infty p_n^{(k)} \log p_n^{(k)}.
\end{equation}
%
In order to prove that $E^{(k)}$ is a monotonically increasing function of $k$, we proceed as follows.
The Pascal identity
\begin{equation}
\binom{n+k+1}{k+1} = \binom{n+k}{k+1} + \binom{n+k}{k},
\end{equation}
allows us to write
\begin{equation}
p_n^{(k+1)} = \lambda^2p_{n-1}^{(k+1)} + (1-\lambda^2)p_n^{(k)},
\end{equation}
where we set $p_{n}^{(k)}=0$ for $n<0$ by definiteness.
Now, using the strict concavity of the function $h[x]=-x\log x$, we have
\begin{equation}
\sum_{n=0}^\infty h\left[p_n^{(k+1)}\right]  > \lambda^2 \sum_{n=0}^\infty h\left[p_{n-1}^{(k+1)}\right] + (1-\lambda^2) \sum_{n=0}^\infty  h\left[p_n^{(k)}\right],
\label{eq:entropic_inequality}
\end{equation}
for $0<\lambda<1$.
Since $p_{n-1}^{(k+1)}$ is equivalent to $p_{n}^{(k+1)}$ up to a shift to the right in the Fock basis,
which does not change the entropy, i.e., $\sum_{n=0}^\infty h\left[p_{n-1}^{(k+1)}\right] = \sum_{n=0}^\infty h\left[p_n^{(k+1)}\right]$,
the expression (\ref{eq:entropic_inequality}) is simply equivalent to
\begin{equation}
E^{(k+1)} > E^{(k)}
\end{equation}
for $0<\lambda<1$. Thus, we conclude that the entanglement can only increase with the number of photon additions or subtractions
when acting on one mode only (before or after the two-mode squeezer).

\begin{figure*}[t]
\includegraphics[width=0.95\textwidth]{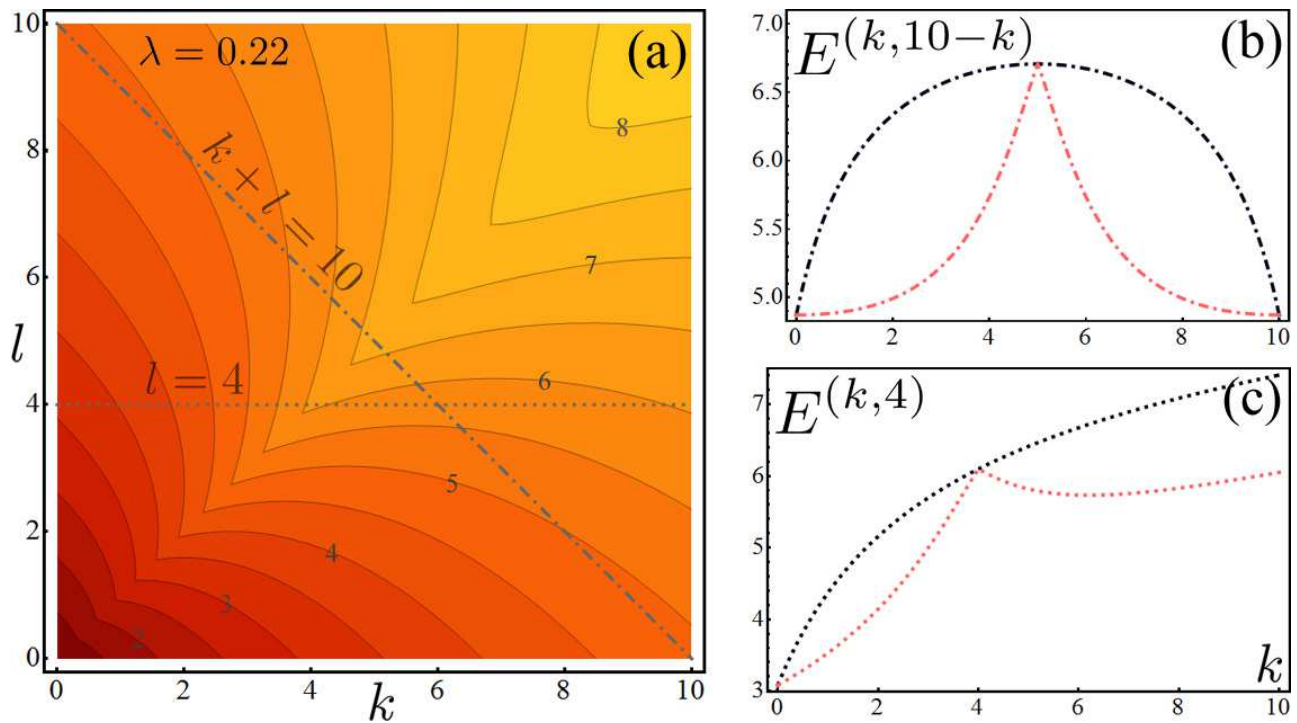}
\caption{\label{Fig4}(Color online) Same as Fig.~\ref{Fig3} but for $\lambda=0.22$. For conciseness, we only show $E_{\mathrm{sub}}^{(k,l)}$ as a density plot (a) as well as the entanglement along the lines $k+l=10$ (b, dashed-dotted line) and  $l=4$ (c, dotted line), both for the photon-added (dark-blue) and photon-subtracted (light-red) states. Note that while the photon-added states have the same behavior as for $\lambda=0.4$, this is not the case for the photon-subtracted states, see details in the text.}
\end{figure*}

\section{Entanglement properties for two-mode operations}

The non-trivial expressions of $p_n^{(k,l)}$ and $q_n^{(k,l)}$ have prevented us from doing an exhaustive analytical study of the entanglement properties of states (\ref{Add}) and (\ref{Sub}) when operating on both modes, that is, when $k\neq0$ and $l\neq0$. Indeed the only interesting property that we have been able to prove analytically is that $E_\mathrm{add}^{(k,k)}=E_\mathrm{sub}^{(k,k)}$, that is, for symmetric operation (same number of operations on both modes $k=l$), additions and subtractions lead to the exact same entanglement, a result already noted in \cite{Dellano07} in the $k=l=1$ case. In order to prove this, note that by renaming the summation index as $n=m+k$, the subtracted state can be written as
\begin{eqnarray}
|\psi_\mathrm{sub}^{(k,k)}\rangle &=& \sum_{m=0}^\infty \frac{\lambda^n\binom{n+k}{k}}{\sqrt{_2F_1(k+1,k+1;1;\lambda^2)}} |m,m\rangle \nonumber\\
&=& \sum_{m=0}^\infty \sqrt{p_m^{(k,k)}} |m,m\rangle,
\end{eqnarray}
which implies that $|\psi_\mathrm{sub}^{(k,k)}\rangle$ and $|\psi_\mathrm{add}^{(k,k)}\rangle$ have the same Schmidt coefficients, hence the same entanglement.

In the reminder of the section, we analyze the entanglement of the states (\ref{Add}) and (\ref{Sub}) by numerically performing the sums (\ref{Eadd}) and (\ref{Esub}), truncated at an upper limit that ensures that the distributions $p_n^{(k,l)}$ and $q_n^{(k,l)}$ are normalized up to an accuracy of $10^{-10}$. Our numerical results are plotted in Fig.~\ref{Fig3} for $\lambda=0.4$ and Fig.~\ref{Fig4} for $\lambda=0.22$.

The main tendency we can deduce from our numerical analysis is that it is always better to perform addition rather than subtraction in order
to increase the entanglement, i.e., $E_\mathrm{add}^{(k,l)} > E_\mathrm{sub}^{(k,l)}$.
This is clearly visible when comparing the density plots in Figs.~\ref{Fig3}a and \ref{Fig3}b. Such a result could be linked to the fact that photon addition seems to increase the state's non-classicality faster than photon subtraction does \cite{Jones97,Kim05}. Note, however, that for large squeezing the difference becomes less pronounced,
i.e., $E_\mathrm{sub}^{(k,l)} \rightarrow E_\mathrm{add}^{(k,l)}$ for $\lambda\rightarrow 1$.

It is also important to remark that the probabilities of success of the addition and subtraction schemes are different \cite{subtraction,Marek08,Zhang10}, and hence, even though addition performs better for the same number of operations, it might be preferable to perform more subtractions to achieve a given entanglement, depending on the particular experimental scenario.

\begin{figure*}[t]
\includegraphics[width=\textwidth]{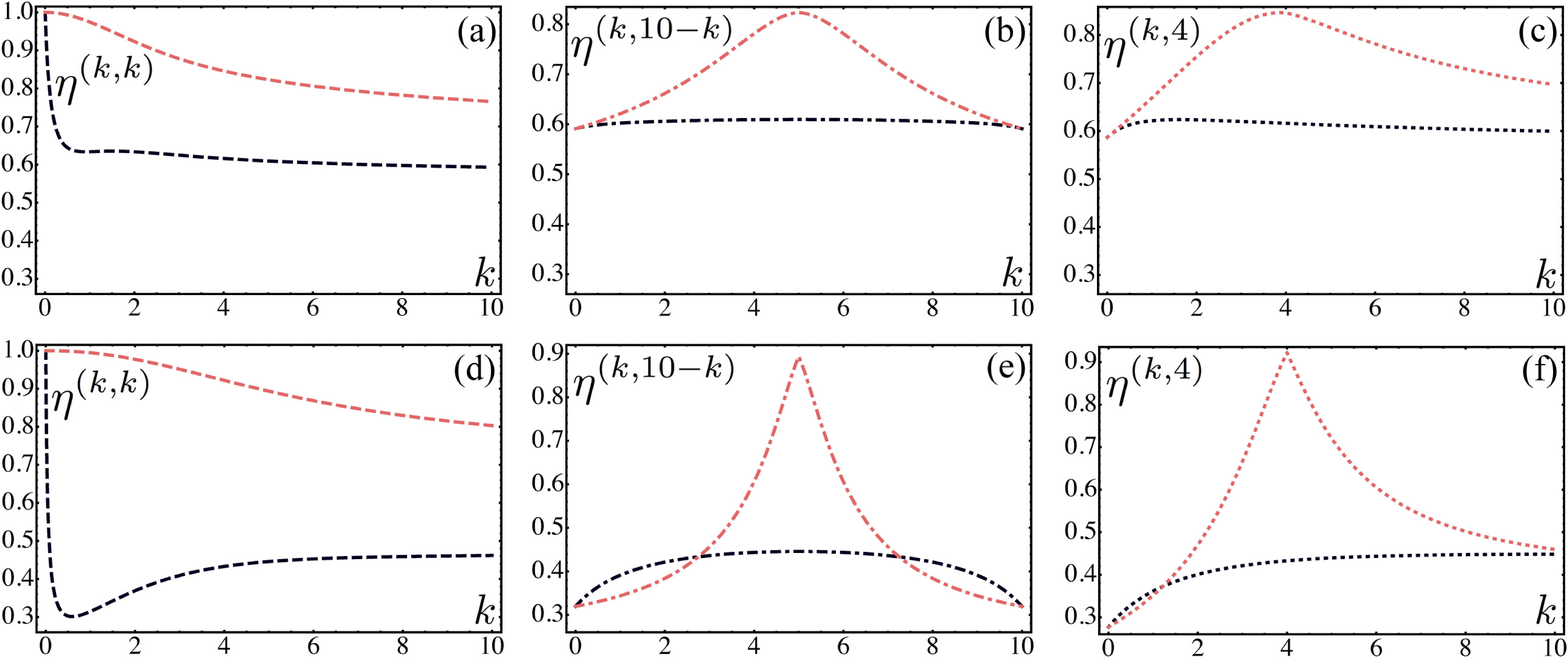}
\caption{\label{FigEff}(Color online) Entanglement energy-efficiency $\eta^{(k,l)}$ of the photon-added (dark-blue curves)
and photon-subtracted (light-red curves) states as a function of the number of operations $k$ and $l$
for $\lambda=0.4$ (upper row) and $\lambda=0.22$ (lower row). As we did for the entanglement,
we show the evolution of the efficiency along the $k=l$ (a,d), $k+l=10$ (b,e), and $l=4$ (c,f) lines.}
\end{figure*}

For symmetric operation $k=l$, where addition and subtraction perform equally,
the entanglement increases with the number of operations,
i.e., $E^{(k+1,k+1)}>E^{(k,k)}$. This is explicitly shown in Fig.~\ref{Fig3}c,
where we plot the entanglement $E^{(k,k)}$ as a function of $k$.
This behavior is in agreement with the studies performed in \cite{Adesso09,Zhang10}.

We also observed that for a fixed number of operations $k+l$, the entanglement increases
when approaching the symmetric situation $k=l$.
This is shown in Fig. \ref{Fig3}d, where we plot $E^{(k,10-k)}$
as a function of $k$ both for addition (dark-blue line) and subtraction (light-red line).
Note, however, that the shapes of the curves are rather different for addition and subtraction.
Nevertheless, we remark that the optimal enhancement is obtained when the same number of operations are applied on both modes,
where both addition and subtraction give the same entanglement enhancement.

By keeping the number of operations fixed on one mode,
the entanglement is an increasing function of the number of additions on the other mode,
that is, $E_\mathrm{add}^{(k+1,l)}  > E_\mathrm{add}^{(k,l)}$.
Thus, by fixing $l$ to some value, say $l_0$, the entanglement increases as
Alice adds more photons; for $l_0 = 0$ this is exactly the analytical result that
we found in Sec.~III. The case $l_0=4$ is illustrated
in Fig.~\ref{Fig3}e. The situation is a bit different for photon subtraction.
While above some critical squeezing parameter $\lambda$ (depending on $l_0$),
the entanglement is a monotonically increasing function of the number of
subtractions $k$ for a fixed $l=l_0$, just as for additions, below this critical squeezing it is not.
This is made clear in Fig.~\ref{Fig4}, where we plot $E_\mathrm{sub}^{(k,l)}$
for a smaller squeezing parameter $\lambda=0.22$. Note Fig.~\ref{Fig4}c, in particular,  where we see
that the entanglement decreases in some interval of $k$ above the symmetric point $k=l=4$
before going back to its normal increase. Otherwise, the behavior of entanglement at $\lambda=0.22$
as shown in Figs.~\ref{Fig4}a and \ref{Fig4}b is qualitatively similar to what we observed at $\lambda=0.4$ in Fig.~\ref{Fig3}.
The case of photon addition is also plotted in Figs.~\ref{Fig4}b and \ref{Fig4}c for comparison.

Comparing Figs.~\ref{Fig3} and \ref{Fig4}, we also observe that
the entanglement enhancement effected by photon addition and subtraction is greater, in relative terms,
when the squeezing parameter $\lambda$ is low (remember that  $E_\mathrm{add}^{(k,l)}$ and $E_\mathrm{sub}^{(k,l)}$
are normalized to the TMVS state in the figures). For example, at the symmetric point $k=l=4$,
the entanglement is enhanced by a factor about 3.7 with respect to the TMVS state at  $\lambda=0.4$,
while it reaches about 6.7 at $\lambda=0.22$.
This can be interpreted as follows. For $\lambda \to 1$, the
entanglement of the TMSV state is already very large and its Schmidt coefficients approach a uniform,
infinitely wide distribution; hence, photon addition or subtraction cannot improve much on this.


Finally, it is worth comparing the photon-added and -subtracted states,
$|\psi_\mathrm{add}^{(k,l)}\rangle$ and $|\psi_\mathrm{sub}^{(k,l)}\rangle$,
from the point of view of the energy cost for generating the same amount of entanglement. For this, we define
the entanglement energy-efficiency of the photon-added state as
\begin{equation}
\eta_\mathrm{add}^{(k,l)}= \frac {E_\mathrm{add}^{(k,l)}} {g\left(N_\mathrm{add}^{(k,l)}/2 \right)}
\end{equation}
where $N_\mathrm{add}^{(k,l)}
= \langle \psi_\mathrm{add}^{(k,l)}  | \hat{a}^\dagger  \hat{a} +\hat{b}^\dagger  \hat{b} |\psi_\mathrm{add}^{(k,l)}\rangle$
is the total mean photon number of the state, and the function $g(x)= (x+1) \log(x+1) - x \log(x)$ is the entanglement entropy of a TMSV state with $2x$ mean photon number (equal to the entropy of a thermal state of $x$ photons). Of course, we use a similar definition for photon-subtracted states.

Taking into account that the TMSV state provides the highest entanglement for a given average photon number, the entanglement energy-efficiency as defined here is equal to one for a TMSV state and less than one otherwise. In other words, the efficiency quantifies the degree to which the state's energy is optimally deployed in creating entanglement.

In Fig. \ref{FigEff}, we plot the entanglement energy-efficiency as a function of the number of operations $(k,l)$, for two values of $\lambda$. Note that, even though photon addition leads to a larger entanglement amplification in absolute terms as shown before, the results shown in Fig. \ref{FigEff} tell us that photon subtraction is more efficient (in general) from the energy cost point of view.

\section{Non-Gaussianity of the photon-added and -subtracted states}

In this section, we evaluate the non-Gaussianity of the photon-added and -subtracted states that we have introduced in the previous sections,
and investigate the possible link with their entanglement properties. In a nutshell, we reach the conclusion that
photon addition leads to a faster degaussification of the TMSV state than photon subtraction, which is reminiscent to the behavior of entanglement,
but nevertheless the level of entanglement found in these states seems to have no direct relation to their non-Gaussianity.

We use here the non-Gaussianity measure of a state that was introduced in \cite{Genoni08}.
This measure has already been used in a similar context, for example in \cite{Allegra10}. There,
after a numerical analysis based on this measure,
it was conjectured that, at least for the class of photon-number entangled states (to which the states included in this work belong),
the entanglement of Gaussian states is more robust against lossy channel with thermal added noise than that of non-Gaussian states.
Note, however, that this conjecture was recently proved wrong by showing that it does not hold when a different entanglement measure (negativity under partial transposition) is used \cite{Lee11}, and that the entanglement of the N00N states and of a simple class of photon-number entangled states
survive longer in a thermal environment than the entanglement of any Gaussian state \cite{Sabapathy11}.

Let us first explain how this non-Gaussianity measure $\bar{G}[\hat{\rho}]$ works
for a general state $\hat{\rho}$  \cite{Genoni08}.
The idea is to evaluate the statistical distinguishability between $\hat{\rho}$ and the Gaussian state
$\hat{\rho}_\mathrm{G}$ having the same first and second moments, which
can be done by using the quantum relative entropy. Thus, we define the non-Gaussianity of state $\hat{\rho}$ as
\begin{eqnarray}
\bar{G}[\hat{\rho}]&=&\mathcal{S}[\hat{\rho}||\hat{\rho}_\mathrm{G}] =
\mathrm{tr}\{\hat{\rho}(\log\hat{\rho}-\log\hat{\rho}_\mathrm{G})\}  \nonumber \\
&=& \mathcal{S}[\hat{\rho}_\mathrm{G}]-\mathcal{S}[\hat{\rho}].
\end{eqnarray}
where the last equality follows from the fact that $\hat{\rho}$ and $\hat{\rho}_\mathrm{G}$
have the same first and second moments.
Here, the states $|\psi_{\mathrm{add}}^{(k,l)}\rangle$ and
$|\psi_{\mathrm{sub}}^{(k,l)}\rangle$ we work with are pure, and hence their non-Gaussianity
is simply the entropy of the corresponding Gaussian state, i.e.,
$\bar{G}[\hat{\rho}]=\mathcal{S}[\hat{\rho}_\mathrm{G}]$.

\begin{figure*}[t]
\includegraphics[width=\textwidth]{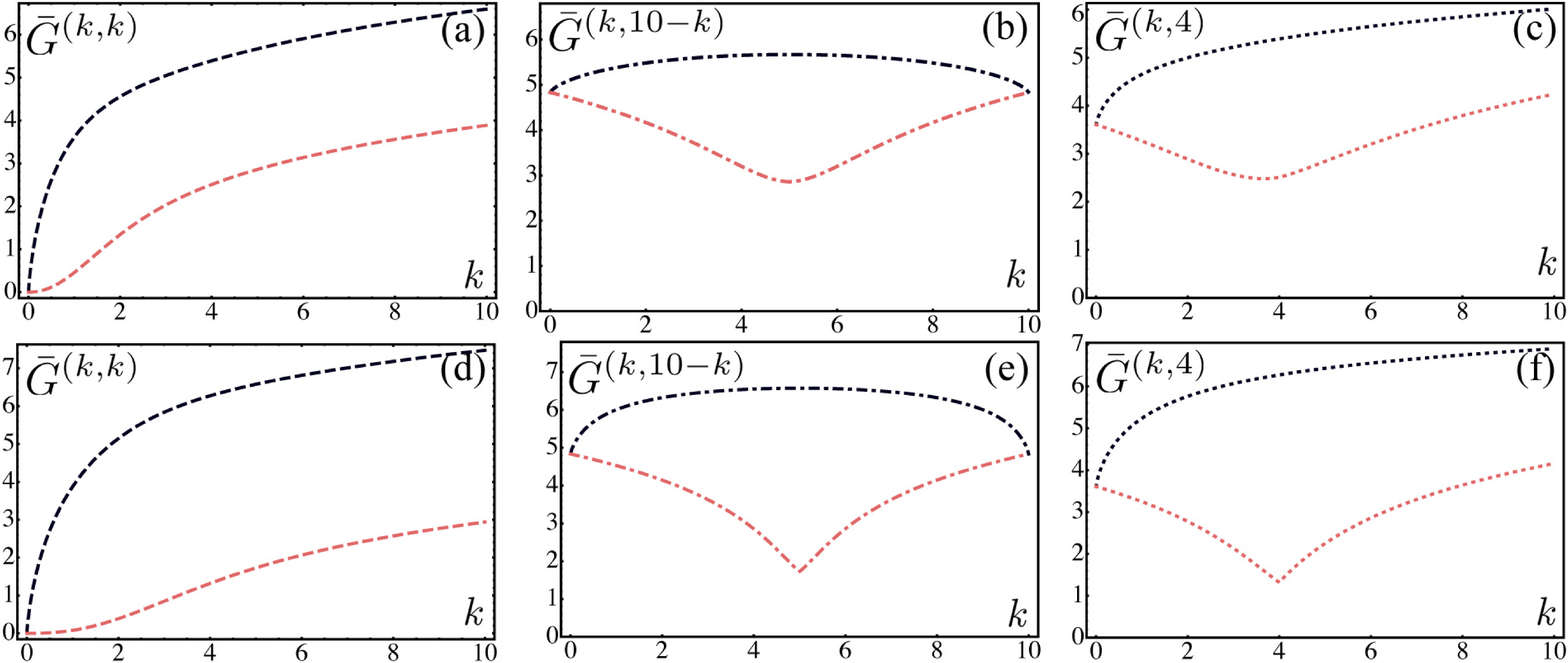}
\caption{\label{Fig5}(Color online) Non-Gaussianity $\bar{G}^{(k,l)}$ of the photon-added (dark-blue curves)
and photon-subtracted (light-red curves) states as a function of the number of operations $k$ and $l$
for $\lambda=0.4$ (upper row) and $\lambda=0.22$ (lower row). As we did for the entanglement and the entanglement energy-efficiency, we show the evolution of the non-Gaussianity along the $k=l$ (a,d), $k+l=10$ (b,e), and $l=4$ (c,f) lines. Note that, in contrast with the entanglement and the entanglement energy-efficiency, the non-Gaussianity has the same qualitative behavior for both $\lambda$ values.}
\end{figure*}

Now, let us define the vector operator $\mathbf{\hat{r}}=(\hat{x}_a,\hat{p}_a,\hat{x}_b,\hat{p}_b)$ built on the quadrature operators
$\hat{x}_a=\hat{a}^\dagger+\hat{a}$ and $\hat{p}_a=i(\hat{a}^\dagger-\hat{a})$, and similarly for the mode \textit{B}. It is fairly simple to check
that the states $|\psi_{\mathrm{add}}^{(k,l)}\rangle$ and $|\psi_{\mathrm{sub}}^{(k,l)}\rangle$ have all zero mean,
that is, $\langle\mathbf{\hat{r}}\rangle=\mathbf{0}$. The elements of the covariance matrix are then evaluated as $C_{jl}=\langle \hat{r}_j \hat{r}_l +\hat{r}_l \hat{r}_j \rangle/2$, and it is straightforward to check that both the photon-subtracted and photon-added states have a covariance matrix
of the type
\begin{equation}\label{CovMat}
C^{(k,l)}=
\left[
\begin{array}{c|c}
\alpha^{(k,l)} \; \openone  & \gamma^{(k,l)} \; \sigma_z\\
\hline
\gamma^{(k,l)} \; \sigma_z & \beta^{(k,l)} \; \openone
\end{array}
\right],
\end{equation}
where we have defined the $2\times2$ matrices $\openone=\mathrm{diag}(1,1)$ and $\sigma_{z}=\mathrm{diag}(1,-1)$. In the case of the photon-added states $|\psi_\mathrm{add}^{(k,l)}\rangle$, the parameters are given by
\begin{eqnarray}
\alpha_\mathrm{add}^{(k,l)} &=& 1+2k+2 \sum_{n=0}^{\infty} np_n^{(k,l)},
\\
\beta_\mathrm{add}^{(k,l)} &=& 1+2l+2 \sum_{n=0}^{\infty} np_n^{(k,l)}, \nonumber
\\
\gamma_\mathrm{add}^{(k,l)} &=& 2 \sum_{n=0}^{\infty} \sqrt{(n+k+1)(n+l+1)p_n^{(k,l)}p_{n+1}^{(k,l)}} \nonumber;
\end{eqnarray}
while, for the photon-subtracted states $|\psi_\mathrm{sub}^{(k,l)}\rangle$, one has
\begin{eqnarray}
\alpha_\mathrm{sub}^{(k,l)} &=& 1-2k+2\sum_{n=\max(k,l)}^\infty nq_n^{(k,l)},
\\
\beta_\mathrm{sub}^{(k,l)} &=& 1-2l+2\sum_{n=\max(k,l)}^\infty nq_n^{(k,l)}, \nonumber
\\
\gamma_\mathrm{sub}^{(k,l)} &=& 2\sum_{n=\max(k,l)}^\infty\sqrt{(n-k+1)(n-l+1)q_n^{(k,l)}q_{n+1}^{(k,l)}} \nonumber.
\end{eqnarray}
While we have not been able to evaluate the sums in the $\gamma$'s analytically, the sums in the $\alpha$'s and $\beta$'s
have the following closed expressions:
\begin{eqnarray}
\sum_{n=0}^\infty n p_n^{(k,l)}  &=& (1+k)(1+l) \lambda^2 \frac{_2F_1(k+2,l+2;2;\lambda^2)}{_2F_1(k+1,l+1;1;\lambda^2)},
\nonumber \\
\sum_{n=k}^\infty n q_n^{(k,l)}  &=& k + (1+k) \lambda^2 \frac{\binom{1+k}{l}}{\binom{k}{l}}
\\
&&\times\frac{_2F_1(k+2,k+2;k-l+2;\lambda^2)}{_2F_1(k+1,k+1;k-l+1;\lambda^2)}, \nonumber
\end{eqnarray}
where in the second expression we have assumed $k \geq l$ (once again, exactly the same expression but interchanging $k$ and $l$ holds for $k<l$).

The two-mode covariance matrix (\ref{CovMat}) is in normal form \cite{WeedbrookUN}, from which the entropy of the Gaussian state (and thus the non-Gaussianity of the photon-added or -subtracted state) can be directly evaluated as
\begin{equation}
\bar{G}^{(k,l)} = g[\nu_+^{(k,l)}] + g[\nu_-^{(k,l)}],
\end{equation}
where
\begin{equation}
g(z)=\frac{z+1}{2} \log \frac{z+1}{2}-\frac{z-1}{2} \log\frac{z-1}{2},
\end{equation}
and where
\begin{equation}
\nu_\pm^{(k,l)} = \left[ \sqrt{ \left( \frac{\alpha^{(k,l)}}{4}+\frac{\beta^{(k,l)}}{4}\right)^{2}-\gamma^{(k,l)2} } \pm \frac{\alpha^{(k,l)}-\beta^{(k,l)}}{2} \right],
\end{equation}
are the symplectic eigenvalues \cite{WeedbrookUN} of the covariance matrix (\ref{CovMat}).

In Fig.~\ref{Fig5}, we plot the non-Gaussianity $\bar{G}^{(k,l)}$ for an arbitrary number of operations $(k,l)$
with $\lambda=0.4$ (upper row) and $\lambda=0.22$ (lower row).
In analogy with the behavior of entanglement, we observe that
photon addition leads to a faster degradation of the Gaussianity of the TMSV state than photon subtraction.
In other words, $|\psi_{\mathrm{add}}^{(k,l)}\rangle$ (dark-blue lines) is more non-Gaussian
than the photon subtracted state $|\psi_{\mathrm{sub}}^{(k,l)}\rangle$ (light-red lines). This is clear for example
in Figs.~\ref{Fig5}a,d where we plot the increase of  $\bar{G}^{(k,k)}$ with $k$ for symmetric operations $k=l$.
Extrapolating from the behavior of entanglement, one would be tempted to predict that
the non-Gaussianity $\bar{G}^{(k,l)}$ is maximum for symmetric operations $k=l$ for both addition and subtraction.
Interestingly, the behavior of $\bar{G}^{(k,l)}$ is radically different.
While for a fixed number $k+l$ of photon additions (dark-blue lines),
its maximum is indeed reached for $k=l$, for a fixed number $k+l$ of photon subtractions (light-red lines),
the non-Gaussianity is actually minimum for $k=l$, see Figs.~\ref{Fig5}b,e.
We observe a similar anomaly in Figs.~\ref{Fig5}c,f, where we plot the non-Gaussianity as a function of $k$ for a fixed $l$.
Thus, the non-Gaussianity of photon-subtracted states exhibits, in some situations, a very different qualitative behavior
from that of its entanglement.


\section{Conclusions}

We have studied how local photon addition and subtraction affect the entanglement and
Gaussianity of the two-mode squeezed vacuum state. This subject has become of recent interest,
especially since these fundamental heralded non-Gaussian operations have become
available in the laboratory.

First, we have analytically shown that the entanglement grows
with the number of photon additions or subtractions when only one of the parties performs the operations.
We have then numerically analyzed the case in which both parties add or subtract photons; although
addition and subtraction lead to the same entanglement enhancement when both parties perform the
same number of operations, photon addition beats photon subtraction in general.

We have also analyzed the efficiency with which the energy in photon-added and photon-subtracted states generates entanglement, showing that, in general, this can be close to perfect for photon subtraction, but not for photon addition.

Finally, we have
numerically studied the degaussification of the two-mode squeezed vacuum state that is effected
by photon addition or subtraction, showing
that photon addition degrades the Gaussianity of the state faster than photon subtraction.
Observing that the entanglement and non-Gaussianity of photon-subtracted states
have a radically different behavior, we conclude that the relation between entanglement
and non-Gaussianity is not as simple as previously assumed.

Future directions might include analyzing how photon addition and subtraction affect the entanglement of more general (possibly mixed) states, such as the TMSV state degraded by losses in the channels through which the entangled modes are sent to Alice and Bob, which is a typical initial state of many distillation or concentration protocols \cite{Zhang10}.
\\
\begin{acknowledgments}
We acknowledge comments and suggestions from David Adjiashvili, G\'eza Giedke, Seth Lloyd, J. Ignacio Cirac, Eugene Polzik, Hyunchul Nha, Myungshik Kim, Eugenio Rold\'an, Germ\'an J. de Valc\'arcel, and Jonas S. Neergaard-Nielsen. C.N.-B. and N.J.C. thank the Optical and Quantum Communications Group at RLE for their hospitality. C.N.-B. acknowledges financial support from the MICINN through the FPU program, from the Spanish Government and the European Union FEDER through Project FIS2008-06024-C03-01, and from the European Comission through project MALICIA under Open grant number: 265522; R.G.-P., N.J.C., and J.H.S. from the W. M. Keck Foundation Center for Extreme Quantum Information Theory; R.G.-P. from the Humboldt foundation; J.H.S. from the ONR Basic Research Challenge Program; and N.J.C. from the F.R.S.-FNRS under project HIPERCOM.
\end{acknowledgments}


\end{document}